\documentclass[pdflatex,sn-nature,twocolumn,iicol]{sn-jnl}

\usepackage{graphicx}%
\usepackage{multirow}%
\usepackage{amsmath,amssymb,amsfonts}%
\usepackage{amsthm}%
\usepackage{mathrsfs}%
\usepackage[title]{appendix}%
\usepackage{xcolor}%
\usepackage{textcomp}%
\usepackage{manyfoot}%
\usepackage{booktabs}%
\usepackage{algorithm}%
\usepackage{algorithmicx}%
\usepackage{algpseudocode}%
\usepackage{listings}%
\usepackage{tabularx}

\newcolumntype{Y}[1]{>{\hsize=#1\hsize\centering\arraybackslash\small}X}

\theoremstyle{thmstyleone}%
\theoremstyle{thmstyletwo}%

\theoremstyle{thmstylethree}%

\unnumbered

\begin{document}

\title[Article Title]{Tapered Fibre Fabry-Pérot Cavities for the Generation of Near-Visible, Phase-locked, and Broadband Kerr Frequency Combs}


\author*[1,2]{\fnm{Matthew} \sur{Macnaughtan}}\email{matthew.macnaughtan@auckland.ac.nz}

\author[1,2]{\fnm{Yiqing} \sur{Xu}}

\author[1,2]{\fnm{Miro} \sur{Erkintalo}}
\author[1,2]{\fnm{St\'ephane} \sur{Coen}}
\author[1,2]{\fnm{Stuart G.} \sur{Murdoch}}

\affil*[1]{\orgdiv{Physics Department}, \orgname{The University of Auckland}, \orgaddress{\city{Auckland}, \country{New Zealand}}}

\affil[2]{\orgname{The Dodd-Walls Centre for Photonic and Quantum Technologies}, \orgaddress{\city{Auckland}, \country{New Zealand}}}


\abstract{We demonstrate the generation of near-visible, broadband Kerr frequency combs using a tapered fibre Fabry-Pérot resonator. By incorporating micron-diameter tapered silica fibre into a pulsed-driven Fabry-Pérot configuration, we exploit geometric dispersion engineering to overcome the strong normal dispersion intrinsic to silica (and nearly all materials) at short wavelengths, achieving net anomalous dispersion in the near-visible regime. The strong mode confinement of the tapered fibre simultaneously enhances the Kerr nonlinearity by over an order of magnitude relative to standard fibre, substantially reducing the threshold drive power for nonlinear structure formation. Through precise control of the cavity fabrication, we experimentally demonstrate and numerically confirm a rich suite of coherent, phase-locked nonlinear structures, including conventional anomalous dispersion cavity solitons, zero-dispersion solitons with spectral bandwidths exceeding 16~THz between 750~nm and 800~nm, as well as switching wave combs and broadband Raman solitons, the latter also reported here for the first time in the near-visible regime. Our results establish the tapered fibre Fabry-Pérot resonator as a versatile platform for near-visible and potentially visible coherent frequency comb generation.}

\maketitle

\section{Introduction}\label{sec1}

When operated under the correct conditions, coherently driven passive Kerr cavities are well known to produce phase-locked frequency combs. In the frequency domain, these combs consist of multiple equidistant spectral lines in which the spacing is governed by the cavity's free spectral range (FSR). Temporally, Kerr frequency combs correspond to ultra-short nonlinear structures that circulate unaltered and indefinitely within the cavity. To date, phase-locked Kerr combs have found applications across a wide range of areas including spectroscopy~\cite{coddington_dual-comb_2016}, ultra-fast ranging~\cite{trocha_ultrafast_2018}, coherent high throughput telecommunications~\cite{pfeifle_coherent_2014,fujii_dissipative_2022}, optical computing~\cite{xu_11_2021,xu_photonic_2020}, and microwave generation~\cite{liu_photonic_2020,weng_turn-key_2024}.

Kerr frequency combs can exist in a variety of states depending on the dispersion regime. In the anomalous dispersion regime, there exists the temporal cavity soliton (CS)~\cite{leo_temporal_2010,herr_temporal_2014}. CSs exist through a double balance between gain and loss, and dispersion and nonlinearity, and can be considered as a single pattern element of modulation instability (MI) sitting atop a low-power homogenous steady-state (HSS)~\cite{tlidi_localized_1994,firth_optical_1996}. Within the normal dispersion regime, there exists switching wave (SW) Kerr combs. Unlike CSs, SW combs form when the intracavity field exists in a state in which both the upper-HSS and lower-HSS are equally favoured~\cite{coen_convection_1999}, thus forming either a stationary dark~\cite{macnaughtan_temporal_2023,xue_mode-locked_2015}, or bright~\cite{lobanov_generation_2015,lobanov_frequency_2015} structure, depending on the driving conditions. Under conditions of synchronous pulsed-pumping, and because the SW comb components form via dispersive wave (DW) phase-matching~\cite{conforti_dispersive_2013,malaguti_dispersive_2014,talla_mbe_coexistence_2020}, the spectra of SW combs can be efficiently tuned simply by adjusting the pump repetition rate~\cite{xu_frequency_2021}. When the effects of stimulated Raman scattering (SRS) are included, SW combs under pulsed-pumping can be tuned towards longer wavelengths, which, provided certain phase- and group-velocity matching conditions are met, leads to the spontaneous formation of Raman solitons (RSs)~\cite{li_ultrashort_2024,xu_frequency_2021}. These structures are exceptionally broadband, and, like SW combs, are inherently stable and deterministic~\cite{li_ultrashort_2024}. When the dispersion is close to zero, Kerr cavities are capable of sustaining zero-dispersion solitons (ZDSs)~\cite{anderson_zero_2022,xiao_near-zero-dispersion_2023,li_experimental_2020}. ZDSs sit at the intersection of anomalous and normal dispersion nonlinear structures, as such they inherit properties of both. For example, ZDSs are known to exhibit higher-order multi-peak pulses and a collapsed snaking bifurcation structure~\cite{li_experimental_2020} analogous to dark-pulse SW combs~\cite{parra-rivas_origin_2016}, as well as bright, structurally independent single peaked CS states~\cite{li_observations_2021}. Spectrally, higher-order ZDSs exhibit DWs on both the upper and lower HSSs, which, due to prominent third-order dispersion, sit asymmetrically around the pump~\cite{talla_mbe_coexistence_2020,jang_observation_2014}. While the nonlinear structures mentioned above each posses different and useful characteristics, generating them within the same type of resonator is typically very difficult due to the carefully optimised dispersion parameters required in each case. 

Furthermore, the vast majority of modern Kerr frequency combs like those mentioned above operate in telecom bands, namely the near-infrared (IR) around 1550~nm. Within this frequency band, there exists a comprehensive industrial base of lasers, electro-optic (EO) devices, optical fibre, and integrated-optical technologies which has enabled rapid progress in both comb quality~\cite{anderson_photonic_2020,li_ultrashort_2024,xiao_near-zero-dispersion_2023,moille_parametrically_2024,moille_kerr-induced_2023,yi_single-mode_2017,xue_super-efficient_2019,xue_microresonator_2017,helgason_surpassing_2023,li_efficiency_2022}, and ease-of-use~\cite{weng_turn-key_2024,voloshin_dynamics_2021,shen_integrated_2020,macnaughtan_soliton_2025}. However, limiting Kerr comb generation to IR wavelengths constrains the technology’s future potential. In particular, at near-visible ($\sim 700-800$~nm) and visible wavelengths ($\sim 400-700$~nm), there exists numerous applications that would benefit from a broadband Kerr comb source such as atomic metrology~\cite{takamoto_optical_2005,ludlow_optical_2015}, astronomical sensing~\cite{benedick_visible_2010,bouchy_fundamental_2001}, and bio-imaging and optical coherence tomography~\cite{xu_toward_2025,wu_visible_2025,lee_ultrahigh_2001}. However, generating useful near-visible and visible Kerr frequency combs has so far proven difficult. The reason for this is primarily due to the dispersion requirements. Nearly all materials that are amenable for resonator fabrication at near-visible and visible wavelengths exhibit strong normal dispersion. For example, within integrated circuits, both bulk silicon nitride and aluminium nitride exhibit second-order dispersion coefficients, $\beta_2$, on the order of 100 $\text{ps}^2/$km at 780 nm~\cite{moss_new_2013,li_aluminium_2021}. Even within silica-based fibre systems, second-order dispersion coefficients around 40 $\text{ps}^2$/km at 780~nm are to be expected~\cite{xu_toward_2025}. Such dispersion figures are not in and of themselves restrictive to the formation of Kerr frequency combs. Indeed, as was mentioned earlier, SW combs can exist under conditions of normal dispersion~\cite{macnaughtan_temporal_2023,lihachev_platicon_2022,xu_frequency_2021,parra-rivas_origin_2016}. However, as the bandwidth of these structures scales with $1/\sqrt{|\beta_2|}$~\cite{coen_universal_2013,leo_dynamics_2013}, any normal dispersion structure that could form would have a correspondingly narrow spectral bandwidth. 

To date, there have been a number of schemes reported to overcome the difficulties outlined above. Earlier efforts focused on utilising both second- and third-order nonlinear processes to translate Kerr combs from the near-IR into the near-visible and visible regimes~\cite{jung_green_2014,guo_efficient_2018}. Although effective, these approaches must contend with the tight phase matching and power requirements which ultimately constrain the total bandwidth and comb power that the secondary comb can inherit. More recently, dispersion engineering schemes within microresonators that manipulate the resonator geometry and mode interactions have been used to generate broadband Kerr combs around 780~nm~\cite{lee_towards_2017,karpov_photonic_2018,liu_near-visible_2025,soltani_enabling_2016,ma_visible_2019,savchenkov_kerr_2011,zhao_broadband_2025}. While promising, these platforms introduce both greater resonator complexity and increased fabrication costs.

In this article, we utilise micron-diameter tapered silica optical fibre within a Fabry-Pérot resonator configuration under synchronous pulsed-pumping to generate anomalous, near-zero, and normal dispersion near-visible Kerr frequency combs. By simply varying the length of untapered fibre during fabrication, we demonstrate the generation of CSs, ZDSs, SW combs, and RSs in the near-visible regime. Notably, the ZDSs produced in this article are, to the best of our knowledge, the broadest passive Kerr-generated phase-locked structures reported in the near-visible regime to date. As an additional advantage, the optical Kerr nonlinearity, $\gamma$, in our TFFP is over an order of magnitude higher than in standard fibre platforms (primarily due to the strong mode confinement), thereby enabling Kerr comb generation at significantly reduced pump powers comparable to that found in microresonators. This work cements TFFPs as not only a versatile platform, but a platform which can effectively offset the strong material dispersion at shorter wavelengths, potentially opening the door for future true visible Kerr comb generation. 

\section{Results}\label{sec2}

\subsection{Experimental Implementation}
\begin{figure}[ht]
    \centering\includegraphics[width=\linewidth]{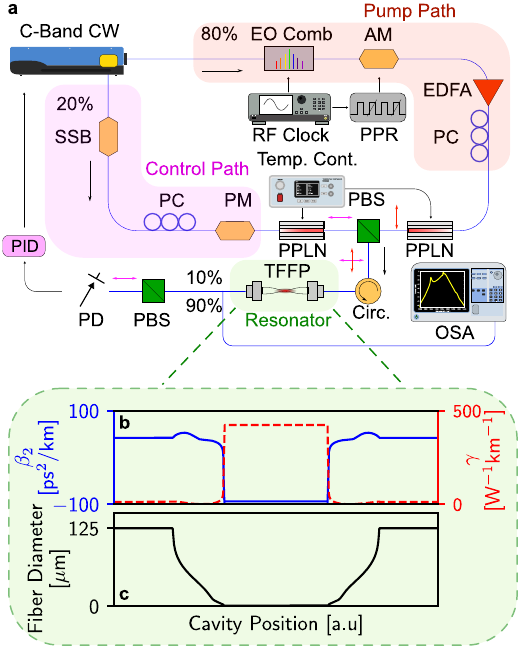}
    \caption{(a) Schematic of our experimental setup. AM: amplitude modulator; PPR: pulse picker; EDFA: erbium-doped fibre amplifier; PC: polarisation controller; PBS: polarisation beam splitter; SSB: single-sideband generator; PM: phase modulator; OSA: optical spectrum analyser; PD: photodiode; PID: proportional-integral-derivative controller; PPLN: periodically poled lithium niobate. (b) Illustration of $\beta_2$ (solid blue curve left axis), and $\gamma$ (dashed red curve right axis) across the TFFP length. (c) Illustration of how the fibre diameter varies across the TFFP length.}
    \label{fig1}
\end{figure}

To experimentally demonstrate near-visible Kerr comb generation, we use the experimental setup shown in Fig.~\ref{fig1}(a), at the centre of which is our TFFP resonator. Our resonator is formed from a single section of single-mode Corning HI-780 fibre with high-reflectivity dielectric mirrors butt coupled to each end. A uniform segment of micron diameter taper [well-known to exhibit both anomalous dispersion~\cite{cordeiro_engineering_2005}, and extremely high Kerr nonlinear coefficients~\cite{magi_enhanced_2007} at near-visible and visible wavelengths as seen in Fig.~\ref{fig1}(b)] is then fabricated (see "Materials and Methods"). Unlike other fibre platforms which may obtain similar dispersion profiles to our TFFP resonators (such as photonic crystal fibres), the addition of untapered fibre [see Fig.~\ref{fig1}(c)] allows for the efficient coupling between the driving and intracavity fields, a task which is typically not possible in resonators with strong mode confinement. Furthermore, the length of untapered fibre can be easily adjusted and used to tailor the resonator's dispersion depending on the type of nonlinear structure that is desired (where the overall dispersion and nonlinearity are averaged over the whole TFFP)~\cite{nielsen_invited_2018,jang_observation_2014}. To demonstrate the precision with which we can control the TFFP dispersion at near-visible wavelengths, we fabricate three different TFFP cavities (labelled as TFFPs A, B, and C) in which the only change is the length of untapered fibre (parameters summarised in "Materials and Methods" Table~\ref{Table1}). 

To drive our TFFP resonators, we generate a train of near-visible (centred at a wavelength $\lambda$ of 770.5~nm) ultra-short pulses (in which the repetition rate is closely matched to the TFFP FSR and controls the pump desynchronisation $\Delta T$) derived from a 1541~nm EO-comb (see "Materials and Methods"). The detuning $\delta$ of our driving field with respect to the nearest cavity resonance is then controlled using a separate control signal (see "Materials and Methods") like the approach outlined in Refs.~\cite{li_ultrashort_2024,macnaughtan_soliton_2025}.
\begin{figure*}[ht]
    \centering\includegraphics[width=\linewidth]{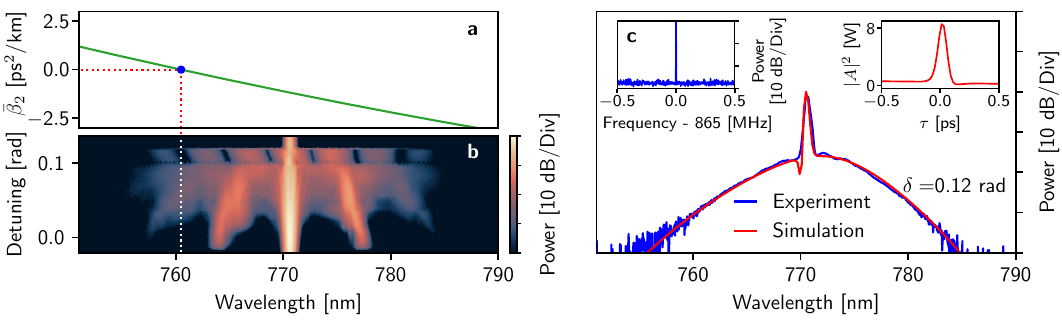}
    \caption{(a) The average second-order dispersion $\bar{\beta}_2$ as a function of wavelength for TFFP A, with the blue dot at 760.5~nm highlighting the ZDW. (b) Pseudo-colour spectral map of a detuning scan at a peak drive power of $P_{\text{p}}$ = 0.75~W. (c) Spectral profile of the intracavity field obtained experimentally (blue curve) and numerically (red curve) at a detuning of 0.12~rad. The left inset shows the experimental electronic beat note (measured from the optical signal filtered between 773~nm and 780~nm) of the CS state, while the right inset shows the temporal intensity profile obtained through simulation.}
    \label{fig2}
\end{figure*}

\subsection{Anomalous Dispersion Cavity Solitons}

We will first demonstrate that our TFFP resonators can indeed fully compensate the material dispersion of the fibre, and stably support coherent, phase-locked anomalous dispersion CS structures in the near-visible regime. To do this, we will use TFFP-A (with an FSR of 865~MHz and a finesse of 110). TFFP-A contains the shortest length of untapered fibre considered in this article, meaning that at the near-visible pump wavelength of 770.5~nm, the anomalous dispersion of the taper waist dominates, leading to an overall anomalous dispersion cavity. Using the dispersion coefficients for TFFP-A seen in Table~\ref{Table1}, we can extrapolate the average second-order dispersion $\bar{\beta}_2$ over a large range of wavelengths [Fig.~\ref{fig2}(a)], and see that the zero-dispersion wavelength (ZDW) of TFFP-A sits around 760.5~nm [blue dot in Fig.~\ref{fig2}(a)]. To test near-visible CS generation, we scan the detuning across a cavity resonance at a peak-pump power of $P_{\text{p}}$ = 0.75~W. As shown in Fig.~\ref{fig2}(b), the emergence of MI sidebands provides a good indicator that the overall dispersion at 770.5~nm is indeed anomalous. At large detunings, the spectral envelope evolves into a modulated pattern (corresponding to CS bound states), and then (for a small range of detunings) into a smooth $\text{sech}^{2}$ profile [shown in blue in Fig.~\ref{fig2}(c)], confirming near-visible single CS formation. Furthermore, by measuring the beat note of the filtered spectral signal (between 773-780~nm), we further confirm in the top left inset of Fig.~\ref{fig2}(c) that the cavity does indeed enter a low-noise, phase-locked CS state. Using the cavity parameters alongside the driving parameters used in Fig.~\ref{fig2}(c), we numerically simulate (see "Materials and Methods") the intracavity field in the CS state [red curve in Fig.~\ref{fig2}(c)], which, when compared to the experimentally acquired data (blue curve), shows excellent agreement.

\subsection{Near-Zero Dispersion Solitons}

\begin{figure}[t]
    \centering\includegraphics[width=\linewidth]{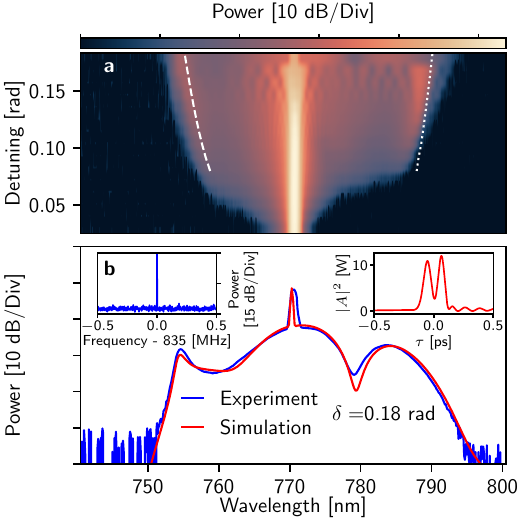}
    \caption{(a) Pseudo-coloured spectral map of a detuning ramp where $P_{\text{p}} = 1$~W in TFFP-B. The dashed (dotted) white curve corresponds to the lower (upper) state DW phase matching solution. (b) Single spectral trace showing both an experimentally acquired 2-peak ZDS (blue curve), and a simulated one (red curve) when $\delta = 0.18$~rad and $P_{\text{p}} = 1$~W. The top right inset shows the corresponding temporal intensity profile obtained via simulation, whereas the top left inset shows the experimental electronic beat note.}
    \label{fig3}
\end{figure}

While we have demonstrated in Fig.~\ref{fig2} with TFFP-A that our platform can indeed fully compensate the natural material dispersion of the fibre to create near-visible anomalous dispersion CSs, as mentioned earlier, a whole suite of useful Kerr frequency combs exist. In particular, ZDSs near the boundary between anomalous and normal dispersion regimes. To illustrate the versatility of our TFFP platform, we will generate near-visible ZDSs using TFFP-B which has an FSR of 835~MHz and finesse of 116. Compared to the previous cavity, TFFP-B has an extra 4.6~mm of untapered fibre present, increasing the overall normal dispersion contribution and causing the ZDW (around 772.1~nm) to move to longer wavelengths (see "Materials and Methods" Table~\ref{Table1} for the dispersion and nonlinear coefficients). Conducting a detuning scan at a peak drive power of $P_{\text{p}} = 1$~W, we obtain the experimental pseudo-coloured spectral plot shown in Fig.~\ref{fig3}(a). Unlike Fig.~\ref{fig2}, we do not see the emergence of MI sidebands, but broadband DWs that sit asymmetrically around the pump, with the dashed (dotted) white curve corresponding to the lower state (upper state) phase-matching condition~\cite{li_experimental_2020,jang_observation_2014,conforti_dispersive_2013}. These features correspond to a higher-order pulse structure maintained through third-order dispersion, which, as the detuning is increased, collapses towards a few-peaked ZDS structure~\cite{parra-rivas_stable_2017,li_experimental_2020}. Figure~\ref{fig3}(b) shows an experimental spectrum (blue curve) of the intracavity field at a detuning of $\delta = 0.18$~rad. Comparing this experimental spectrum to simulation under the same parameters (red curve), we see good agreement between the two. Looking at the temporal intensity profile obtained through simulation [right inset in Fig.~\ref{fig3}(b)], we can see that the structure present corresponds to a near-visible 2-peak ZDS in which each peak maintains a FWHM of $\approx70$~fs and is separated from the other by $\approx110$~fs. By measuring the electronic beat note of the filtered (between 773~nm and 780~nm) spectral components [left inset in Fig.~\ref{fig3}(b)], we further confirm that this structure is indeed a coherent, phase-locked state of the intracavity field. Moreover, with a spectral bandwidth exceeding 16 THz, the ZDS shown in Fig.~\ref{fig3}(b) is, to the best of our knowledge, the spectrally broadest soliton structure yet demonstrated in a passive Kerr cavity pumped in the near-visible regime.

\subsection{Switching Waves and Raman Solitons}
\begin{figure*}[ht]
    \centering\includegraphics[width=\linewidth]{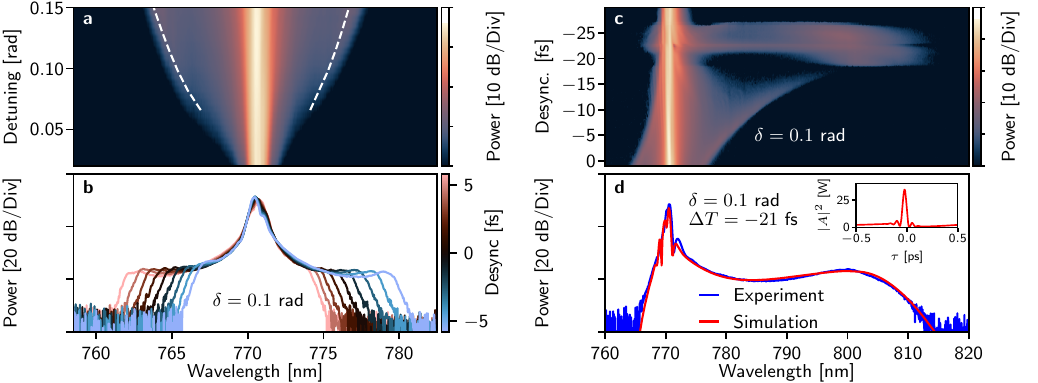}
    \caption{(a) Pseudo-coloured spectral map showing a detuning ramp in TFFP-C with $P_{p} = 1.2$~W and $\Delta T = 0$~fs, where the white dashed curves represent the lower state DW phase matching solutions. (b) Single spectral traces of the intracavity field (at $\delta = 0.1$~rad and $P_{p} = 1.2$~W) as $\Delta T$ is swept from positive to negative values (where varying trace colour correspond to different $\Delta T$). (c) Pseudo-coloured spectral map showing a desynchronisation ramp with $P_{p} = 1.2$~W and $\delta = 0.1$~rad. (d) Single spectral trace of the intracavity field when $P_{p} = 1.2$~W, $\Delta T = -21$~fs, and $\delta = 0.1$~rad. The blue (red) curve is the experimentally (numerically) acquired spectra, with the inset showing the temporal intensity profile obtained through simulation.}
    \label{fig6}
\end{figure*}

Next, to further demonstrate the versatility and generality of our TFFP platform, we show that even conventional normal dispersion SW combs and RSs can be tailored for in the near-visible regime. To generate broader near-visible SW combs, we use TFFP-C with an FSR and finesse of 803~MHz and 134, respectively. As with the previous cavity, TFFP-C has an added 9~mm of untapered fibre, leading to a greater normal dispersion contribution and a longer ZDW around 795.1~nm (see "Materials and Methods" Table~\ref{Table1} for the dispersion and nonlinear coefficients). Performing a detuning scan at zero pump desynchronisation, Fig.~\ref{fig6}(a) shows the emergence of symmetric DWs that move progressively further from the pump as the detuning increases. Such behaviour is characteristic of normal-dispersion SW combs, with the experimentally observed DW frequencies agreeing well with the phase-matched frequencies predicted from the calculated dispersion and nonlinear coefficients [dashed white curves in Fig.~\ref{fig6}(a)]~\cite{talla_mbe_coexistence_2020,macnaughtan_temporal_2023}. Furthermore, by varying the pump desynchronisation, Fig.~\ref{fig6}(b) demonstrates that the DWs can be selectively shifted towards both longer wavelengths (negative desynchronisations) and shorter wavelengths (positive desynchronisations). As the desynchronisation is driven to increasingly negative values, Fig.~\ref{fig6}(c) further reveals the sudden formation of a broadband near-visible structure at a desynchronisation of $\Delta T = -21$~fs [blue curve in Fig.~\ref{fig6}(d)]. Numerical simulations performed using the same driving parameters [red curve in Fig.~\ref{fig6}(d)] show strong agreement with experiment and identify the resulting structure as an ultrashort, phase-locked RS~\cite{li_ultrashort_2024,xu_frequency_2021}. The near-visible RS shown in Fig.~\ref{fig6}(d) is not only exceptionally broad (the spectrum shown in Fig.~\ref{fig6}(d) spans over 16~THz), but is the first reported RS generated at near-visible wavelengths. 

\section{Discussion}\label{sec12}

In this work, we have shown both experimentally and numerically that by incorporating a tapered fibre section into a pulsed-driven Fabry–Pérot configuration, the resonator dispersion at near-visible wavelengths can not only be pushed into the anomalous dispersion regime, but simply and actively engineered (Fig.~\ref{fig1}), thus enabling the generation of the full range of near-visible coherent, phase-locked bright structures known to manifest themselves in the anomalous, near-zero, and normal dispersion regimes. By simply varying the length of untapered fibre within the cavity, we show that the ZDW can be accurately shifted around the 770.5~nm pump wavelength, facilitating the formation of conventional anomalous dispersion CSs (Fig.~\ref{fig2}), ZDSs (Fig.~\ref{fig3}), as well as SW combs and RSs (Fig.~\ref{fig6}). Moreover, owing to the strong mode confinement, the optical Kerr nonlinearity exceeds that of conventional fibre cavities by more than an order of magnitude, substantially reducing the threshold pump powers for nonlinear structure formation into the sub Watt range. 

Interestingly, as seen in Figs.~\ref{fig2}-\ref{fig6}, the nonlinear structures we generate maintain extremely clean spectra; that is, we see no evidence of Kelly sidebands~\cite{nielsen_invited_2018,smith_sideband_1992} or avoided-mode crossings~\cite{lucas_detuning-dependent_2017}, as is commonly seen in dispersion-engineered (and managed) resonators~\cite{li_experimental_2020,nielsen_invited_2018}. In the case of Kelly sidebands, this is somewhat surprising because dispersion management is known to increase the efficiency of sideband formation by orders of magnitude~\cite{nielsen_invited_2018}. However, unlike traditional fibre-based dispersion-managed cavities, the adiabatic evolution of the guided mode plays a central role in our platform~\cite{nagai_ultra-low-loss_2014}, meaning that the guided mode is not shedding energy at discrete transition points within the cavity (thus amplifying Kelly sidebands)~\cite{bednyakova_adiabatic_2015}. Regarding avoided-mode crossings, our system is inherently protected against them because we exclusively use the fundamental guided mode (see "Materials and Methods"). While the thinnest part of our TFFP can support some higher-order modes (with a normalised frequency of $V = 3.5$), the fundamental mode always maintains a large index difference from the higher-order modes, preventing coupling. Moreover, because we use single-mode fibre (at our pump wavelength) to construct our TFFP cavities, the untapered fibre present within our TFFPs further prevents the emergence of any higher-order modes.

Concerning future research avenues, the generation of true visible passive Kerr solitons is of particular interest. For example, considering a pump wavelength near 650~nm, preliminary simulations (not shown) suggest that a taper waist diameter of 0.75~$\mu$m and a taper length of 60~mm are sufficient to achieve net anomalous dispersion in the 630–650~nm range (assuming 3~cm of untapered fibre within the cavity). In the visible regime, the principal challenge is currently not the engineering of the required cavity dispersion, but rather the availability of a suitable pump source capable of driving the resonator. Nevertheless, as demonstrated in this work, nonlinear frequency conversion techniques, such as sum-frequency generation, offer a viable route towards producing stable, narrow-linewidth, and tunable visible pump sources. As such, the TFFP platform stands as a compelling candidate for the first realisation of passive Kerr solitons in the visible spectrum.

\section{Materials and Methods}\label{sec11}
\begin{table*}[ht]
\centering
\begin{tabularx}{1.0\linewidth}{ |Y{0.5}||Y{1.1}|Y{1.0}|Y{1.1}|Y{1.1}|Y{1.2}| }
 \hline
 TFFP & Untapered Length [cm] & $\bar{\beta}_2$ [$\text{ps}^{2}/\text{km}$]& $\bar{\beta}_3$ [$\text{ps}^{3}/\text{km}$]& $\bar{\beta}_4$ [$\text{ps}^{4}/\text{km}$]& $\bar{\gamma}$ [1/$\text{W}/\text{km}$]\\
 \hline
 A & 4.44 & $-$1.2 & 3.5$\times10^{-2}$ & 5.9$\times10^{-5}$ & 146\\
 B & 4.90 & 0.2 & 3.6$\times10^{-2}$ & 6.1$\times10^{-5}$ & 141\\
 C & 5.80 & 2.7 & 3.8$\times10^{-2}$ & 4.6$\times10^{-5}$ & 130\\
 \hline
\end{tabularx}
\caption{Table showing the length of untapered fibre, average $n$-th order dispersion $\bar{\beta}_n$, and average Kerr nonlinearity $\bar{\gamma}$ for the three TFFP cavities considered in this article (labelled A, B, and C).}
\label{Table1}
\end{table*}
\subsection{TFFP Fabrication}

The fabrication process for TFFP resonators begins by attaching ferrules to each end of a small segment of fibre (forming an untapered Fabry-Pérot). The length of this initial segment is an important control parameter and determines the amount of untapered fibre that will be present within the completed TFFP resonators (i.e a longer initial segment will lead to more untapered fibre within the final TFFP). Once each end of the initial fibre segment is polished and butt-coupled to high reflectivity dielectric mirrors (with transmission coefficients, $\theta$, of 0.8\%), we attach each end of the initial untapered Fabry-Pérot to motorised translation stages for the tapering process. To fabricate the taper, we utilise the commonly used flame-brush technique~\cite{brambilla_ultra-low-loss_2004,yao_ultra-long_2020,hoffman_ultrahigh_2014,lee_fabrication_2019} alongside a length optimisation protocol which reduces the length of the taper transition regions [see Fig.~\ref{fig1}(c)]~\cite{nagai_ultra-low-loss_2014}. For each TFFP seen in this article, we used the same taper profile in which the taper diameter and waist length were set to $0.84~\mu$m and 38~mm, respectively. Once the tapering process has been completed, we fix each end of the TFFP with metal rods which not only keeps the taper taut, but allows us to easily transport it. Table~\ref{Table1} shows the average dispersion, nonlinearity, and untapered fibre lengths for each TFFP considered in this article (TFFP A,B, and C).

\subsection{Generating the Driving Pulse-train}

To drive our TFFP resonators, we use a synchronously driven pulsed-pump derived from a 1541~nm CW laser [see the highlighted red area in Fig.~\ref{fig1}(a)]. Using standard EO methods, we generate a picosecond pulse train in the C-band with a tunable repetition rate $f_{\text{rep}}$ matched closely to an integer multiple of the cavity FSR~\cite{macnaughtan_soliton_2025,xu_frequency_2021,obrzud_temporal_2017}. We then use an amplitude modulator (AM) to pulse-pick our pump so that the repetition rate is closely matched to the cavity FSR. Next, we amplify the pulse-picked pump signal [using an erbium doped fibre amplifier (EDFA)] and pass it through a temperature controlled periodically poled lithium niobate (PPLN) crystal for conversion to near-visible wavelengths via SHG. After frequency conversion, we obtain a pulse-train of 2.3~ps pulses with a centre wavelength of 770.5~nm, whose repetition rate is set by the electronic signal generator that drives the original 1541~nm pulse-train.

\subsection{Detuning Control}

We control our detuning using the approach outlined in Refs.~\cite{li_ultrashort_2024,macnaughtan_soliton_2025}. A fraction of the inital CW laser output is diverted to a tunable single sideband (SSB) generator, passed through a phase modulator (PM) for Pound-Drever-Hall locking~\cite{nielsen_invited_2018}, and frequency-converted to 770.5~nm via PPLN. This control signal is then combined with the pump using a polarisation beam-splitter (PBS), ready to be injected into the cavity. At the cavity output, a second PBS directs the control polarisation towards a slow photodiode (PD), whose signal feeds a proportional-integral-derivative (PID) controller that locks the control signal on resonance. Adjusting the frequency applied to the SSB modulator then directly tunes the pump detuning. 

\subsection{Numerically Modelling the TFFP Intracavity Field}

To numerically model the intracavity field in a way that accurately captures the dispersive and nonlinear management of our cavity, we use the following infinite-dimensional Ikeda map~\cite{coen_passive_1999,coen_experimental_1998,haelterman_dissipative_1992,agrawal_nonlinear_2013}:
\begin{multline}\label{Boundary_conditions}
    A^{(m+1)}(0,\tau) = (1-2\alpha) A^{(m)}(L,\tau)\text{exp}(-i\delta) \\ + \sqrt{\theta}A_{\text{in}}(\tau),
\end{multline}
\begin{multline}\label{NLSE}
    \frac{\partial A^{(m)}(z,\tau)}{\partial z} = \Biggl( \frac{\Delta T}{L}\frac{\partial }{\partial \tau} + i \sum_{n=2}\frac{i^{n}\beta_{n}(z)}{n!}\frac{\partial^{n} }{\partial \tau^{n}} \\
    + i\gamma(z)\Bigl( (1-f_{\text{R}})|A^{(m)}|^{2} + f_{\text{R}}h_{\text{R}}\ast|A^{(m)}|^{2} \Bigr) \Biggr)A^{(m)},
\end{multline}
where $A^{(m)}(z,\tau)$ represents the intracavity field envelope at the $m-$th roundtrip, $\tau$ represents the fast-time and describes how the intracavity field evolves in a reference frame travelling at the group velocity of the driving field, and $z$ represents the distance along the cavity. Equation~(\ref{Boundary_conditions}) acts as the boundary conditions and includes our coherent driving $A_{\text{in}}(\tau)$, linear detuning from the nearest resonance $\delta$, and loss term $\alpha$. The driving term is related to the pump power, $P(\tau)$, by $P(\tau) = |A_{\text{in}}(\tau)|^2$, and is modelled numerically by a Gaussian temporal profile matching that of the pump pulses used in our experiment as $A_{\text{in}}(\tau) = \sqrt{P_{\text{p}}}\text{exp}(-2.77\tau^{2}/T_{0}^2)$ where $P_{\text{p}}$ and $T_0$ represent the peak pump-power and pulse full width at half-maximum (FWHM), respectively. Equation~(\ref{NLSE}) is the well known nonlinear Schrödinger equation (NLSE) and describes the evolution of the intracavity field within a single roundtrip. The first two terms describe the pump desynchronisation from the intracavity field (with $\Delta T = f_{\text{rep}}^{-1} - \text{FSR}^{-1}$ being our desynchronisation term) and the chromatic dispersion (truncated at 4th order). The final terms model the nonlinearities and includes both the instantaneous Kerr response and delayed Raman response $h_{\text{R}}$ (which we model using a multivibrational model~\cite{hollenbeck_multiple-vibrational-mode_2002}) with Raman fraction $f_{\text{R}} = 0.18$. 

\subsection{Numerically Modelling the Fibre Dispersion and Nonlinearity}

To model how the dispersion and nonlinearity change throughout our TFFP, we consider the evolution of the fundamental $HE_{11}$ mode (higher-order modes are not relevant due to the usage of single-mode step-index fibre). For larger fibre diameters, we model the guided mode using the standard two-layer approach~\cite{snyder_optical_1983}. As the fibre diameter decreases, we eventually take into account the influence of the surrounding medium (i.e. air) and consider a three-layer model~\cite{erdogan_cladding-mode_1997}. By solving the appropriate eigenvalue equations, we are able to obtain the propagation constant $\beta$ for a range of fibre diameters and wavelengths. With $\beta$, we are then able to obtain the mode profile, from which we can calculate the Kerr nonlinearity $\gamma$. The dispersion coefficients follow directly from the propagation constant as $\beta_n = \text{d}^n\beta/\text{d}\omega^n$ (where $\omega$ is the angular frequency). Figure~\ref{fig1}(b) gives an example of how $\beta_2$ and $\gamma$ vary across a TFFP. 

\backmatter

\bmhead{Acknowledgements}

The Authors would like to thank Simon Barter at the University of Auckland for their help in imaging our samples, and Miles Anderson at the University of Auckland for fruitful discussions.

\section*{Declarations}
\bmhead{Disclosures}
The authors declare no conflicts of interest.
\bmhead{Data availability} Data underlying the results presented in this paper are not publicly available at this time but may be obtained from the authors upon reasonable request.
\bmhead{Code availability} The code that supports the plots within this paper and other findings of this study are available from the corresponding author upon reasonable request.
\bmhead{Funding} Marsden fund of the Royal Society of New Zealand.
\bmhead{Author Contributions} M.M performed all experimental work alongside assistance from Y.X. Numerical simulations were completed by M.M with the help of S.G.M. The first draft of the article was written by M.M, with editing and reviews done by S.G.M, M.E, and S.C. The overall theory and concept was developed by Y.X, S.G.M, and M.M. Overall supervision of the project was provided by S.G.M, with additional support and supervision given by M.E and S.C.


\bibliography{references}

@book{agrawal_nonlinear_2013,
	edition = {Fifth edition},
	title = {Nonlinear {Fiber} {Optics}},
	author = {Agrawal, Govind P.},
	year = {2013},
}

@article{lee_ultrahigh_2001,
	title = {Ultrahigh scanning speed optical coherence tomography using optical frequency comb generators},
	volume = {40},
	url = {https://iopscience.iop.org/article/10.1143/JJAP.40.L878/meta},
	doi = {10.1143/JJAP.40.L878},
	language = {en},
	number = {8B},
	urldate = {2026-04-27},
	journal = {Japanese Journal of Applied Physics},
	publisher = {IOP Publishing},
	author = {Lee, Seok-Jeong and Widiyatmoko, Bambang and Kourogi, Motonobu and Ohtsu, Motoichi},
	month = aug,
	year = {2001},
}

@article{parra-rivas_stable_2017,
	title = {Stable dark and bright soliton {Kerr} combs can coexist in normal dispersion resonators},
	volume = {95},
	issn = {2469-9926, 2469-9934},
	url = {http://arxiv.org/abs/1609.08819},
	doi = {10.1103/PhysRevA.95.053863},
	language = {en},
	number = {5},
	urldate = {2024-06-10},
	journal = {Physical Review A},
	author = {Parra-Rivas, P. and Gomila, D. and Gelens, L.},
	month = may,
	year = {2017},
	pages = {053863},
}

@phdthesis{coen_passive_1999,
	address = {Bruxelles, Belgium},
	type = {Ph.{D}. dissertation},
	title = {Passive {Nonlinear} {Optical} {Fiber} {Resonators} {Fundamentals} and {Applications}},
	school = {Université libre de Bruxelles},
	author = {Coen, Stephane},
	year = {1999},
}

@article{wu_visible_2025,
	title = {Visible light optical coherence tomography: technology and biomedical applications},
	volume = {12},
	copyright = {http://creativecommons.org/licenses/by/3.0/},
	issn = {2306-5354},
	shorttitle = {Visible {Light} {Optical} {Coherence} {Tomography}},
	url = {https://www.mdpi.com/2306-5354/12/7/770},
	doi = {10.3390/bioengineering12070770},
	language = {en},
	number = {7},
	urldate = {2026-04-27},
	journal = {Bioengineering},
	publisher = {Multidisciplinary Digital Publishing Institute},
	author = {Wu, Songzhi and Wang, Shuo and Li, Baihan and Wang, Zhao},
	month = jul,
	year = {2025},
	pages = {770},
}

@article{soltani_enabling_2016,
	title = {Enabling arbitrary wavelength frequency combs on chip},
	volume = {10},
	copyright = {© 2015 by WILEY-VCH Verlag GmbH \& Co. KGaA, Weinheim},
	issn = {1863-8899},
	url = {https://onlinelibrary.wiley.com/doi/abs/10.1002/lpor.201500226},
	doi = {10.1002/lpor.201500226},
	number = {1},
	urldate = {2026-04-28},
	journal = {Laser \& Photonics Reviews},
	author = {Soltani, Mohammad and Matsko, Andrey and Maleki, Lute},
	year = {2016},
	pages = {158--162},
}

@article{li_aluminium_2021,
	title = {Aluminium nitride integrated photonics: a review},
	volume = {10},
	copyright = {© 2021 The Authors},
	issn = {2192-8614},
	shorttitle = {Aluminium nitride integrated photonics},
	url = {https://onlinelibrary.wiley.com/doi/abs/10.1515/nanoph-2021-0130},
	doi = {10.1515/nanoph-2021-0130},
	language = {en},
	number = {9},
	urldate = {2026-06-24},
	journal = {Nanophotonics},
	author = {Li, Nanxi and Ho, Chong Pei and Zhu, Shiyang and Fu, Yuan Hsing and Zhu, Yao and Lee, Lennon Yao Ting},
	year = {2021},
	pages = {2347--2387},
}

@article{bouchy_fundamental_2001,
	title = {Fundamental photon noise limit to radial velocity measurements},
	volume = {374},
	copyright = {© ESO, 2001},
	issn = {0004-6361},
	url = {https://www.aanda.org/articles/aa/abs/2001/29/aa1316/aa1316.html},
	language = {en-gb},
	number = {2},
	urldate = {2026-04-28},
	journal = {Astronomy \& Astrophysics},
	author = {Bouchy, F. and Pepe, F. and Queloz, D.},
	year = {2001},
	pages = {733--739},
}

@article{xu_photonic_2020,
	title = {Photonic {Perceptron} {Based} on a {Kerr} {Microcomb} for {High}-{Speed}, {Scalable}, {Optical} {Neural} {Networks}},
	volume = {14},
	copyright = {© 2020 Wiley-VCH GmbH},
	issn = {1863-8899},
	url = {https://onlinelibrary.wiley.com/doi/abs/10.1002/lpor.202000070},
	doi = {10.1002/lpor.202000070},
	language = {en},
	number = {10},
	urldate = {2026-06-22},
	journal = {Laser \& Photonics Reviews},
	author = {Xu, Xingyuan and Tan, Mengxi and Corcoran, Bill and Wu, Jiayang and Nguyen, Thach G. and Boes, Andreas and Chu, Sai T. and Little, Brent E. and Morandotti, Roberto and Mitchell, Arnan and Hicks, Damien G. and Moss, David J.},
	year = {2020},
	pages = {2000070},
}

@article{voloshin_dynamics_2021,
	title = {Dynamics of soliton self-injection locking in optical microresonators},
	volume = {12},
	issn = {2041-1723},
	url = {https://www.nature.com/articles/s41467-020-20196-y},
	doi = {10.1038/s41467-020-20196-y},
	language = {en},
	number = {1},
	urldate = {2024-06-13},
	journal = {Nature Communications},
	author = {Voloshin, Andrey S. and Kondratiev, Nikita M. and Lihachev, Grigory V. and Liu, Junqiu and Lobanov, Valery E. and Dmitriev, Nikita Yu. and Weng, Wenle and Kippenberg, Tobias J. and Bilenko, Igor A.},
	month = jan,
	year = {2021},
	pages = {235},
}

@article{helgason_surpassing_2023,
	title = {Surpassing the nonlinear conversion efficiency of soliton microcombs},
	volume = {17},
	issn = {1749-4893},
	url = {https://www.nature.com/articles/s41566-023-01280-3},
	doi = {10.1038/s41566-023-01280-3},
	language = {en},
	number = {11},
	urldate = {2024-06-12},
	journal = {Nature Photonics},
	author = {Helgason, Oskar and Girardi, Marcello and Ye, Zhichao and Lei, Fuchuan and Schröder, Jochen and Torres-Company, Victor},
	month = nov,
	year = {2023},
	pages = {992--999},
}

@article{anderson_photonic_2020,
	title = {Photonic chip-based resonant supercontinuum},
	volume = {8},
	url = {http://arxiv.org/abs/1909.00022},
	doi = {10.1364/OPTICA.403302},
	language = {en},
	number = {6},
	urldate = {2024-06-10},
	journal = {Optica},
	author = {Anderson, Miles H. and Bouchand, Romain and Liu, Junqiu and Weng, Wenle and Obrzud, Ewelina and Herr, Tobias and Kippenberg, Tobias J.},
	month = mar,
	year = {2020},
	pages = {771--779},
}

@article{jang_observation_2014,
	title = {Observation of dispersive wave emission by temporal cavity solitons},
	volume = {39},
	issn = {0146-9592, 1539-4794},
	url = {http://arxiv.org/abs/1406.2046},
	doi = {10.1364/OL.39.005503},
	language = {en},
	number = {19},
	urldate = {2024-06-10},
	journal = {Optics Letters},
	author = {Jang, Jae K. and Erkintalo, Miro and Murdoch, Stuart G. and Coen, Stephane},
	month = oct,
	year = {2014},
	pages = {5503},
}

@article{parra-rivas_origin_2016,
	title = {Origin and stability of dark pulse {Kerr} combs in normal dispersion resonators},
	copyright = {© 2016 Optical Society of America},
	url = {https://opg.optica.org/ol/abstract.cfm?uri=ol-41-11-2402},
	doi = {10.1364/OL.41.002402},
	language = {EN},
	urldate = {2026-06-24},
	journal = {Optics Letters, Vol. 41, Issue 11, pp. 2402-2405},
	publisher = {Optica Publishing Group},
	author = {Parra-Rivas, Pedro and Gomila, Damià and Knobloch, Edgar and Coen, Stéphane and Gelens, Lendert},
	month = jun,
	year = {2016},
}

@article{lobanov_generation_2015,
	title = {Generation of platicons and frequency combs in optical microresonators with normal {GVD} by modulated pump},
	volume = {112},
	copyright = {http://iopscience.iop.org/info/page/text-and-data-mining},
	issn = {0295-5075, 1286-4854},
	url = {https://iopscience.iop.org/article/10.1209/0295-5075/112/54008},
	doi = {10.1209/0295-5075/112/54008},
	language = {en},
	number = {5},
	urldate = {2024-06-10},
	journal = {Europhysics Letters},
	author = {Lobanov, Valery E. and Lihachev, Grigory and Gorodetsky, Michael L.},
	month = dec,
	year = {2015},
	pages = {54008},
}

@article{liu_photonic_2020,
	title = {Photonic microwave generation in the {X}- and {K}-band using integrated soliton microcombs},
	volume = {14},
	issn = {1749-4893},
	url = {https://www.nature.com/articles/s41566-020-0617-x},
	doi = {10.1038/s41566-020-0617-x},
	language = {en},
	number = {8},
	urldate = {2024-06-11},
	journal = {Nature Photonics},
	author = {Liu, Junqiu and Lucas, Erwan and Raja, Arslan S. and He, Jijun and Riemensberger, Johann and Wang, Rui Ning and Karpov, Maxim and Guo, Hairun and Bouchand, Romain and Kippenberg, Tobias J.},
	month = aug,
	year = {2020},
	pages = {486--491},
}

@article{xu_11_2021,
	title = {11 {TOPS} photonic convolutional accelerator for optical neural networks},
	volume = {589},
	issn = {1476-4687},
	url = {https://www.nature.com/articles/s41586-020-03063-0},
	doi = {10.1038/s41586-020-03063-0},
	language = {en},
	number = {7840},
	urldate = {2024-06-11},
	journal = {Nature},
	author = {Xu, Xingyuan and Tan, Mengxi and Corcoran, Bill and Wu, Jiayang and Boes, Andreas and Nguyen, Thach G. and Chu, Sai T. and Little, Brent E. and Hicks, Damien G. and Morandotti, Roberto and Mitchell, Arnan and Moss, David J.},
	month = jan,
	year = {2021},
	pages = {44--51},
}

@article{pfeifle_coherent_2014,
	title = {Coherent terabit communications with microresonator {Kerr} frequency combs},
	volume = {8},
	issn = {1749-4893},
	url = {https://www.nature.com/articles/nphoton.2014.57},
	doi = {10.1038/nphoton.2014.57},
	language = {en},
	number = {5},
	urldate = {2024-06-11},
	journal = {Nature Photonics},
	author = {Pfeifle, Joerg and Brasch, Victor and Lauermann, Matthias and Yu, Yimin and Wegner, Daniel and Herr, Tobias and Hartinger, Klaus and Schindler, Philipp and Li, Jingshi and Hillerkuss, David and Schmogrow, Rene and Weimann, Claudius and Holzwarth, Ronald and Freude, Wolfgang and Leuthold, Juerg and Kippenberg, Tobias J. and Koos, Christian},
	month = may,
	year = {2014},
	pages = {375--380},
}

@article{anderson_zero_2022,
	title = {Zero dispersion {Kerr} solitons in optical microresonators},
	volume = {13},
	copyright = {2022 The Author(s)},
	issn = {2041-1723},
	url = {https://www.nature.com/articles/s41467-022-31916-x},
	doi = {10.1038/s41467-022-31916-x},
	language = {en},
	number = {1},
	urldate = {2026-06-21},
	journal = {Nature Communications},
	publisher = {Nature Publishing Group},
	author = {Anderson, Miles H. and Weng, Wenle and Lihachev, Grigory and Tikan, Alexey and Liu, Junqiu and Kippenberg, Tobias J.},
	month = aug,
	year = {2022},
	pages = {4764},
}

@article{magi_enhanced_2007,
	title = {Enhanced {Kerr} nonlinearity in sub-wavelength diameter {As2Se3} chalcogenide fiber tapers},
	copyright = {© 2007 Optical Society of America},
	url = {https://opg.optica.org/oe/abstract.cfm?uri=oe-15-16-10324},
	doi = {10.1364/OE.15.010324},
	language = {EN},
	urldate = {2026-06-17},
	journal = {Optics Express, Vol. 15, Issue 16, pp. 10324-10329},
	publisher = {Optica Publishing Group},
	author = {Mägi, E. C. and Fu, L. B. and Nguyen, H. C. and Lamont, M. R. E. and Yeom, D. I. and Eggleton, B. J.},
	month = aug,
	year = {2007},
}

@article{fujii_dissipative_2022,
	title = {Dissipative {Kerr} soliton microcombs for {FEC}-free optical communications over 100 channels},
	copyright = {© 2022 Optica Publishing Group},
	url = {https://opg.optica.org/oe/abstract.cfm?uri=oe-30-2-1351},
	doi = {10.1364/OE.447712},
	language = {EN},
	urldate = {2026-06-02},
	journal = {Optics Express, Vol. 30, Issue 2, pp. 1351-1364},
	publisher = {Optica Publishing Group},
	author = {Fujii, Shun and Tanaka, Shuya and Ohtsuka, Tamiki and Kogure, Soma and Wada, Koshiro and Kumazaki, Hajime and Tasaka, Shun and Hashimoto, Yosuke and Kobayashi, Yuta and Araki, Tomohiro and Furusawa, Kentaro and Sekine, Norihiko and Kawanishi, Satoki and Tanabe, Takasumi},
	month = jan,
	year = {2022},
}

@article{cordeiro_engineering_2005,
	title = {Engineering the dispersion of tapered fibers for supercontinuum generation with a 1064 nm pump laser},
	copyright = {© 2005 Optical Society of America},
	url = {https://opg.optica.org/ol/abstract.cfm?uri=ol-30-15-1980},
	doi = {10.1364/OL.30.001980},
	language = {EN},
	urldate = {2026-06-01},
	journal = {Optics Letters, Vol. 30, Issue 15, pp. 1980-1982},
	publisher = {Optica Publishing Group},
	author = {Cordeiro, C. M. B. and Wadsworth, W. J. and Birks, T. A. and Russell, P. St J.},
	month = aug,
	year = {2005},
}

@article{smith_sideband_1992,
	title = {Sideband generation through perturbations to the average soliton model},
	volume = {10},
	issn = {1558-2213},
	url = {https://ieeexplore.ieee.org/document/166771},
	doi = {10.1109/50.166771},
	number = {10},
	urldate = {2026-05-26},
	journal = {Journal of Lightwave Technology},
	author = {Smith, N.J. and Blow, K.J. and Andonovic, I.},
	month = oct,
	year = {1992},
	pages = {1329--1333},
}

@article{bednyakova_adiabatic_2015,
	title = {Adiabatic {Soliton} {Laser}},
	volume = {114},
	doi = {10.1103/PhysRevLett.114.113901},
	number = {11},
	journal = {Physical Review Letters},
	author = {Bednyakova, Anastasia},
	year = {2015},
}

@book{snyder_optical_1983,
	address = {London},
	title = {Optical {Waveguide} {Theory}},
	publisher = {Chapman and Hall},
	author = {Snyder, Allan W. and Love, John D.},
	year = {1983},
}

@article{ma_visible_2019,
	title = {Visible {Kerr} comb generation in a high-{Q} silica microdisk resonator with a large wedge angle},
	copyright = {© 2019 Chinese Laser Press},
	url = {https://opg.optica.org/prj/abstract.cfm?uri=prj-7-5-573},
	doi = {10.1364/PRJ.7.000573},
	language = {EN},
	urldate = {2026-04-29},
	journal = {Photonics Research, Vol. 7, Issue 5, pp. 573-578},
	publisher = {Optica Publishing Group},
	author = {Ma, Jiyang and Xiao, Longfu and Gu, Jiaxin and Li, Hao and Cheng, Xinyu and He, Guangqiang and Jiang, Xiaoshun and Xiao, Min},
	month = may,
	year = {2019},
}

@article{savchenkov_kerr_2011,
	title = {Kerr combs with selectable central frequency},
	volume = {5},
	copyright = {2011 Springer Nature Limited},
	issn = {1749-4893},
	url = {https://www.nature.com/articles/nphoton.2011.50},
	doi = {10.1038/nphoton.2011.50},
	language = {en},
	number = {5},
	urldate = {2026-04-29},
	journal = {Nature Photonics},
	publisher = {Nature Publishing Group},
	author = {Savchenkov, A. A. and Matsko, A. B. and Liang, W. and Ilchenko, V. S. and Seidel, D. and Maleki, L.},
	month = may,
	year = {2011},
	pages = {293--296},
}

@article{karpov_photonic_2018,
	title = {Photonic chip-based soliton frequency combs covering the biological imaging window},
	volume = {9},
	copyright = {2018 The Author(s)},
	issn = {2041-1723},
	url = {https://www.nature.com/articles/s41467-018-03471-x},
	doi = {10.1038/s41467-018-03471-x},
	language = {en},
	number = {1},
	urldate = {2026-04-28},
	journal = {Nature Communications},
	publisher = {Nature Publishing Group},
	author = {Karpov, Maxim and Pfeiffer, Martin H. P. and Liu, Junqiu and Lukashchuk, Anton and Kippenberg, Tobias J.},
	month = mar,
	year = {2018},
	pages = {1146},
}

@article{guo_efficient_2018,
	title = {Efficient {Generation} of a {Near}-visible {Frequency} {Comb} via {Cherenkov}-like {Radiation} from a {Kerr} {Microcomb}},
	volume = {10},
	issn = {2331-7019},
	url = {https://link.aps.org/doi/10.1103/PhysRevApplied.10.014012},
	doi = {10.1103/PhysRevApplied.10.014012},
	language = {en},
	number = {1},
	urldate = {2026-04-28},
	journal = {Physical Review Applied},
	author = {Guo, Xiang and Zou, Chang-Ling and Jung, Hojoong and Gong, Zheng and Bruch, Alexander and Jiang, Liang and Tang, Hong X.},
	month = jul,
	year = {2018},
	pages = {014012},
}

@article{jung_green_2014,
	title = {Green, red, and {IR} frequency comb line generation from single {IR} pump in {AlN} microring resonator},
	copyright = {© 2014 Optical Society of America},
	url = {https://opg.optica.org/optica/abstract.cfm?uri=optica-1-6-396},
	doi = {10.1364/OPTICA.1.000396},
	language = {EN},
	urldate = {2026-04-28},
	journal = {Optica, Vol. 1, Issue 6, pp. 396-399},
	publisher = {Optica Publishing Group},
	author = {Jung, Hojoong and Stoll, Rebecca and Guo, Xiang and Fischer, Debra and Tang, Hong X.},
	month = dec,
	year = {2014},
}

@article{moss_new_2013,
	title = {New {CMOS}-compatible platforms based on silicon nitride and {Hydex} for nonlinear optics},
	volume = {7},
	copyright = {2013 Springer Nature Limited},
	issn = {1749-4893},
	url = {https://www.nature.com/articles/nphoton.2013.183},
	doi = {10.1038/nphoton.2013.183},
	language = {en},
	number = {8},
	urldate = {2026-04-28},
	journal = {Nature Photonics},
	publisher = {Nature Publishing Group},
	author = {Moss, David J. and Morandotti, Roberto and Gaeta, Alexander L. and Lipson, Michal},
	month = aug,
	year = {2013},
	pages = {597--607},
}

@article{xu_toward_2025,
	title = {Toward visible ultrafast imaging with a synchronously pumped switching wave {Kerr} frequency comb},
	copyright = {© 2025 Optica Publishing Group},
	url = {https://opg.optica.org/oe/abstract.cfm?uri=oe-33-3-4714},
	doi = {10.1364/OE.551627},
	language = {EN},
	urldate = {2026-04-28},
	journal = {Optics Express, Vol. 33, Issue 3, pp. 4714-4724},
	publisher = {Optica Publishing Group},
	author = {Xu, Yiqing and Coen, Stéphane and Erkintalo, Miro and Murdoch, Stuart G.},
	month = feb,
	year = {2025},
}

@article{benedick_visible_2010,
	title = {Visible wavelength astro-comb},
	copyright = {© 2010 OSA},
	url = {https://opg.optica.org/oe/abstract.cfm?uri=oe-18-18-19175},
	doi = {10.1364/OE.18.019175},
	language = {EN},
	urldate = {2026-04-28},
	journal = {Optics Express, Vol. 18, Issue 18, pp. 19175-19184},
	publisher = {Optica Publishing Group},
	author = {Benedick, Andrew J. and Chang, Guoqing and Birge, Jonathan R. and Chen, Li-Jin and Glenday, Alexander G. and Li, Chih-Hao and Phillips, David F. and Szentgyorgyi, Andrew and Korzennik, Sylvain and Furesz, Gabor and Walsworth, Ronald L. and Kärtner, Franz X.},
	month = aug,
	year = {2010},
}

@article{takamoto_optical_2005,
	title = {An optical lattice clock},
	volume = {435},
	copyright = {2005 Macmillan Magazines Ltd.},
	issn = {1476-4687},
	url = {https://www.nature.com/articles/nature03541},
	doi = {10.1038/nature03541},
	language = {en},
	number = {7040},
	urldate = {2026-04-28},
	journal = {Nature},
	publisher = {Nature Publishing Group},
	author = {Takamoto, Masao and Hong, Feng-Lei and Higashi, Ryoichi and Katori, Hidetoshi},
	month = may,
	year = {2005},
	pages = {321--324},
}

@article{shen_integrated_2020,
	title = {Integrated turnkey soliton microcombs},
	volume = {582},
	copyright = {2020 The Author(s), under exclusive licence to Springer Nature Limited},
	issn = {1476-4687},
	url = {https://www.nature.com/articles/s41586-020-2358-x},
	doi = {10.1038/s41586-020-2358-x},
	language = {en},
	number = {7812},
	urldate = {2026-04-27},
	journal = {Nature},
	publisher = {Nature Publishing Group},
	author = {Shen, Boqiang and Chang, Lin and Liu, Junqiu and Wang, Heming and Yang, Qi-Fan and Xiang, Chao and Wang, Rui Ning and He, Jijun and Liu, Tianyi and Xie, Weiqiang and Guo, Joel and Kinghorn, David and Wu, Lue and Ji, Qing-Xin and Kippenberg, Tobias J. and Vahala, Kerry and Bowers, John E.},
	month = jun,
	year = {2020},
	pages = {365--369},
}

@article{macnaughtan_soliton_2025,
	title = {Soliton self-excitation under pulsed driving in a {Kerr} resonator},
	volume = {7},
	url = {https://link.aps.org/doi/10.1103/v9wr-dv9y},
	doi = {10.1103/v9wr-dv9y},
	number = {4},
	urldate = {2026-04-27},
	journal = {Physical Review Research},
	publisher = {American Physical Society},
	author = {Macnaughtan, Matthew and Li, Zongda and Xu, Yiqing and Wei, Xiaoming and Yang, Zhongmin and Coen, Stéphane and Erkintalo, Miro and Murdoch, Stuart G.},
	month = dec,
	year = {2025},
	pages = {043250},
}

@article{moille_parametrically_2024,
	title = {Parametrically driven pure-{Kerr} temporal solitons in a chip-integrated microcavity},
	volume = {18},
	copyright = {2024 The Author(s), under exclusive licence to Springer Nature Limited},
	issn = {1749-4893},
	url = {https://www.nature.com/articles/s41566-024-01401-6},
	doi = {10.1038/s41566-024-01401-6},
	language = {en},
	number = {6},
	urldate = {2026-04-27},
	journal = {Nature Photonics},
	publisher = {Nature Publishing Group},
	author = {Moille, Grégory and Leonhardt, Miriam and Paligora, David and Englebert, Nicolas and Leo, François and Fatome, Julien and Srinivasan, Kartik and Erkintalo, Miro},
	month = jun,
	year = {2024},
	pages = {617--624},
}

@article{ludlow_optical_2015,
	title = {Optical atomic clocks},
	volume = {87},
	url = {https://link.aps.org/doi/10.1103/RevModPhys.87.637},
	doi = {10.1103/RevModPhys.87.637},
	number = {2},
	urldate = {2026-04-27},
	journal = {Reviews of Modern Physics},
	publisher = {American Physical Society},
	author = {Ludlow, Andrew D. and Boyd, Martin M. and Ye, Jun and Peik, E. and Schmidt, P. O.},
	month = jun,
	year = {2015},
	pages = {637--701},
}

@article{liu_near-visible_2025,
	title = {Near-visible integrated soliton microcombs with detectable repetition rates},
	volume = {16},
	copyright = {2025 The Author(s)},
	issn = {2041-1723},
	url = {https://www.nature.com/articles/s41467-025-60157-x},
	doi = {10.1038/s41467-025-60157-x},
	language = {en},
	number = {1},
	urldate = {2026-04-27},
	journal = {Nature Communications},
	publisher = {Nature Publishing Group},
	author = {Liu, Peng and Ji, Qing-Xin and Liu, Jin-Yu and Ge, Jinhao and Li, Mingxiao and Guo, Joel and Jin, Warren and Gao, Maodong and Yu, Yan and Feshali, Avi and Paniccia, Mario and Bowers, John E. and Vahala, Kerry J.},
	month = may,
	year = {2025},
	pages = {4780},
}

@article{brambilla_ultra-low-loss_2004,
	title = {Ultra-low-loss optical fiber nanotapers},
	copyright = {© 2004 Optical Society of America},
	url = {https://opg.optica.org/oe/abstract.cfm?uri=oe-12-10-2258},
	doi = {10.1364/OPEX.12.002258},
	language = {EN},
	urldate = {2026-02-12},
	journal = {Optics Express, Vol. 12, Issue 10, pp. 2258-2263},
	publisher = {Optica Publishing Group},
	author = {Brambilla, Gilberto and Finazzi, Vittoria and Richardson, David J.},
	month = may,
	year = {2004},
}

@article{lee_fabrication_2019,
	title = {Fabrication method for ultra-long optical micro/nano-fibers},
	volume = {19},
	issn = {1567-1739},
	url = {https://www.sciencedirect.com/science/article/pii/S1567173919302160},
	doi = {10.1016/j.cap.2019.08.018},
	number = {12},
	urldate = {2026-02-09},
	journal = {Current Applied Physics},
	author = {Lee, Donghwa and Lee, Kwang Jo and Kim, Jin-Hun and Park, Kyungdeuk and Lee, Dongjin and Kim, Yoon-Ho and Shin, Heedeuk},
	month = dec,
	year = {2019},
	pages = {1334--1337},
}

@article{yao_ultra-long_2020,
	title = {Ultra-{Long} {Subwavelength} {Micro}/{Nanofibers} {With} {Low} {Loss}},
	volume = {32},
	issn = {1941-0174},
	url = {https://ieeexplore.ieee.org/abstract/document/9146924},
	doi = {10.1109/LPT.2020.3011719},
	number = {17},
	urldate = {2026-02-09},
	journal = {IEEE Photonics Technology Letters},
	author = {Yao, Ni and Linghu, Shuangyi and Xu, Yingxin and Zhu, Runlin and Zhou, Ning and Gu, Fuxing and Zhang, Lei and Fang, Wei and Ding, Wei and Tong, Limin},
	month = sep,
	year = {2020},
	pages = {1069--1072},
}

@article{moille_kerr-induced_2023,
	title = {Kerr-induced synchronization of a cavity soliton to an optical reference},
	volume = {624},
	copyright = {2023 This is a U.S. Government work and not under copyright protection in the US; foreign copyright protection may apply},
	issn = {1476-4687},
	url = {https://www.nature.com/articles/s41586-023-06730-0},
	doi = {10.1038/s41586-023-06730-0},
	language = {en},
	number = {7991},
	urldate = {2025-09-19},
	journal = {Nature},
	publisher = {Nature Publishing Group},
	author = {Moille, Grégory and Stone, Jordan and Chojnacky, Michal and Shrestha, Rahul and Javid, Usman A. and Menyuk, Curtis and Srinivasan, Kartik},
	month = dec,
	year = {2023},
	pages = {267--274},
}

@article{zhao_broadband_2025,
	title = {Broadband {Near}-{Visible} {Frequency} {Comb} {Generation} via {High}-{Order} {Mode} {Dispersion} {Engineering} in a {Microbubble} {Resonator}},
	url = {https://opg.optica.org/jlt/abstract.cfm?uri=jlt-43-5-2226},
	language = {EN},
	urldate = {2025-06-12},
	journal = {Journal of Lightwave Technology, Vol. 43, Issue 5, pp. 2226-2231},
	publisher = {IEEE},
	author = {Zhao, Zhenlin and Dong, Ruiji and Hu, Ya and Wu, Tianhe and Feng, Ziyao and Huang, Zheng and Huang, Zufang and Li, Ming and Zou, Chang-Ling and Sun, Xiankai and Lu, Qijing},
	month = mar,
	year = {2025},
}

@article{lee_towards_2017,
	title = {Towards visible soliton microcomb generation},
	volume = {8},
	copyright = {2017 The Author(s)},
	issn = {2041-1723},
	url = {https://www.nature.com/articles/s41467-017-01473-9},
	doi = {10.1038/s41467-017-01473-9},
	language = {en},
	number = {1},
	urldate = {2025-06-12},
	journal = {Nature Communications},
	publisher = {Nature Publishing Group},
	author = {Lee, Seung Hoon and Oh, Dong Yoon and Yang, Qi-Fan and Shen, Boqiang and Wang, Heming and Yang, Ki Youl and Lai, Yu-Hung and Yi, Xu and Li, Xinbai and Vahala, Kerry},
	month = nov,
	year = {2017},
	pages = {1295},
}

@article{lucas_detuning-dependent_2017,
	title = {Detuning-dependent properties and dispersion-induced instabilities of temporal dissipative {Kerr} solitons in optical microresonators},
	volume = {95},
	url = {https://link.aps.org/doi/10.1103/PhysRevA.95.043822},
	doi = {10.1103/PhysRevA.95.043822},
	number = {4},
	urldate = {2024-10-18},
	journal = {Physical Review A},
	publisher = {American Physical Society},
	author = {Lucas, Erwan and Guo, Hairun and Jost, John D. and Karpov, Maxim and Kippenberg, Tobias J.},
	month = apr,
	year = {2017},
	pages = {043822},
}

@article{yi_single-mode_2017,
	title = {Single-mode dispersive waves and soliton microcomb dynamics},
	volume = {8},
	copyright = {2017 The Author(s)},
	issn = {2041-1723},
	url = {https://www.nature.com/articles/ncomms14869},
	doi = {10.1038/ncomms14869},
	language = {en},
	number = {1},
	urldate = {2024-10-08},
	journal = {Nature Communications},
	publisher = {Nature Publishing Group},
	author = {Yi, Xu and Yang, Qi-Fan and Zhang, Xueyue and Yang, Ki Youl and Li, Xinbai and Vahala, Kerry},
	month = mar,
	year = {2017},
	pages = {14869},
}

@article{xiao_near-zero-dispersion_2023,
	title = {Near-zero-dispersion soliton and broadband modulational instability {Kerr} microcombs in anomalous dispersion},
	volume = {12},
	copyright = {2023 The Author(s)},
	issn = {2047-7538},
	url = {https://www.nature.com/articles/s41377-023-01076-8},
	doi = {10.1038/s41377-023-01076-8},
	language = {en},
	number = {1},
	urldate = {2024-08-19},
	journal = {Light: Science \& Applications},
	publisher = {Nature Publishing Group},
	author = {Xiao, Zeyu and Li, Tieying and Cai, Minglu and Zhang, Hongyi and Huang, Yi and Li, Chao and Yao, Baicheng and Wu, Kan and Chen, Jianping},
	month = feb,
	year = {2023},
	pages = {33},
}

@article{haelterman_dissipative_1992,
	title = {Dissipative modulation instability in a nonlinear dispersive ring cavity},
	volume = {91},
	issn = {0030-4018},
	url = {https://www.sciencedirect.com/science/article/pii/003040189290367Z},
	doi = {10.1016/0030-4018(92)90367-Z},
	number = {5},
	urldate = {2024-06-24},
	journal = {Optics Communications},
	author = {Haelterman, M. and Trillo, S. and Wabnitz, S.},
	month = aug,
	year = {1992},
	pages = {401--407},
}

@article{macnaughtan_temporal_2023,
	title = {Temporal characteristics of stationary switching waves in a normal dispersion pulsed-pump fiber cavity},
	copyright = {© 2023 Optica Publishing Group},
	url = {https://opg.optica.org/ol/abstract.cfm?uri=ol-48-15-4097},
	doi = {10.1364/OL.492998},
	language = {EN},
	urldate = {2024-06-22},
	journal = {Optics Letters, Vol. 48, Issue 15, pp. 4097-4100},
	publisher = {Optica Publishing Group},
	author = {Macnaughtan, Matthew and Erkintalo, Miro and Coen, Stéphane and Murdoch, Stuart and Xu, Yiqing},
	month = aug,
	year = {2023},
}

@article{firth_optical_1996,
	title = {Optical {Bullet} {Holes}: {Robust} {Controllable} {Localized} {States} of a {Nonlinear} {Cavity}},
	volume = {76},
	shorttitle = {Optical {Bullet} {Holes}},
	url = {https://link.aps.org/doi/10.1103/PhysRevLett.76.1623},
	doi = {10.1103/PhysRevLett.76.1623},
	number = {10},
	urldate = {2024-06-13},
	journal = {Physical Review Letters},
	publisher = {American Physical Society},
	author = {Firth, W. J. and Scroggie, A. J.},
	month = mar,
	year = {1996},
	pages = {1623--1626},
}

@article{tlidi_localized_1994,
	title = {Localized structures and localized patterns in optical bistability},
	volume = {73},
	url = {https://link.aps.org/doi/10.1103/PhysRevLett.73.640},
	doi = {10.1103/PhysRevLett.73.640},
	number = {5},
	urldate = {2024-06-13},
	journal = {Physical Review Letters},
	publisher = {American Physical Society},
	author = {Tlidi, M. and Mandel, Paul and Lefever, R.},
	month = aug,
	year = {1994},
	pages = {640--643},
}

@article{coen_universal_2013,
	title = {Universal scaling laws of {Kerr} frequency combs},
	volume = {38},
	url = {https://opg-optica-org.ezproxy.auckland.ac.nz/ol/fulltext.cfm?uri=ol-38-11-1790&id=253664},
	doi = {10.1364/OL.38.001790},
	number = {11},
	journal = {Optics Letters},
	author = {Coen, Stéphane and Erkintalo, Miro},
	year = {2013},
	pages = {1790--1792},
}

@article{weng_turn-key_2024,
	title = {Turn-key {Kerr} soliton generation and tunable microwave synthesizer in dual-mode {Si} $_{\textrm{3}}$ {N} $_{\textrm{4}}$ microresonators},
	volume = {32},
	issn = {1094-4087},
	url = {https://opg.optica.org/abstract.cfm?URI=oe-32-3-3123},
	doi = {10.1364/OE.510228},
	language = {en},
	number = {3},
	urldate = {2024-06-13},
	journal = {Optics Express},
	author = {Weng, Haizhong and McDermott, Michael and Afridi, Adnan Ali and Tu, Huilan and Lu, Qiaoyin and Guo, Weihua and Donegan, John F.},
	month = jan,
	year = {2024},
	pages = {3123},
}

@article{nielsen_invited_2018,
	title = {Invited {Article}: {Emission} of intense resonant radiation by dispersion-managed {Kerr} cavity solitons},
	volume = {3},
	issn = {2378-0967},
	shorttitle = {Invited {Article}},
	url = {https://doi.org/10.1063/1.5060123},
	doi = {10.1063/1.5060123},
	number = {12},
	urldate = {2024-06-13},
	journal = {APL Photonics},
	author = {Nielsen, Alexander U. and Garbin, Bruno and Coen, Stéphane and Murdoch, Stuart G. and Erkintalo, Miro},
	month = nov,
	year = {2018},
	pages = {120804},
}

@article{trocha_ultrafast_2018,
	title = {Ultrafast optical ranging using microresonator soliton frequency combs},
	volume = {359},
	url = {https://www.science.org/doi/10.1126/science.aao3924},
	doi = {10.1126/science.aao3924},
	number = {6378},
	urldate = {2024-06-11},
	journal = {Science},
	publisher = {American Association for the Advancement of Science},
	author = {Trocha, P. and Karpov, M. and Ganin, D. and Pfeiffer, M. H. P. and Kordts, A. and Wolf, S. and Krockenberger, J. and Marin-Palomo, P. and Weimann, C. and Randel, S. and Freude, W. and Kippenberg, T. J. and Koos, C.},
	month = feb,
	year = {2018},
	pages = {887--891},
}

@article{lihachev_platicon_2022,
	title = {Platicon microcomb generation using laser self-injection locking},
	volume = {13},
	issn = {2041-1723},
	url = {https://www.nature.com/articles/s41467-022-29431-0},
	doi = {10.1038/s41467-022-29431-0},
	language = {en},
	number = {1},
	urldate = {2024-06-10},
	journal = {Nature Communications},
	author = {Lihachev, Grigory and Weng, Wenle and Liu, Junqiu and Chang, Lin and Guo, Joel and He, Jijun and Wang, Rui Ning and Anderson, Miles H. and Liu, Yang and Bowers, John E. and Kippenberg, Tobias J.},
	month = apr,
	year = {2022},
	pages = {1771},
}

@article{nagai_ultra-low-loss_2014,
	title = {Ultra-low-loss tapered optical fibers with minimal lengths},
	volume = {22},
	issn = {1094-4087},
	url = {https://opg.optica.org/oe/abstract.cfm?uri=oe-22-23-28427},
	doi = {10.1364/OE.22.028427},
	language = {en},
	number = {23},
	urldate = {2024-06-10},
	journal = {Optics Express},
	author = {Nagai, Ryutaro and Aoki, Takao},
	month = nov,
	year = {2014},
	pages = {28427},
}

@article{obrzud_temporal_2017,
	title = {Temporal solitons in microresonators driven by optical pulses},
	volume = {11},
	issn = {1749-4885, 1749-4893},
	url = {https://www.nature.com/articles/nphoton.2017.140},
	doi = {10.1038/nphoton.2017.140},
	language = {en},
	number = {9},
	urldate = {2024-06-10},
	journal = {Nature Photonics},
	author = {Obrzud, Ewelina and Lecomte, Steve and Herr, Tobias},
	month = sep,
	year = {2017},
	pages = {600--607},
}

@article{li_observations_2021,
	title = {Observations of existence and instability dynamics of near-zero-dispersion temporal {Kerr} cavity solitons},
	volume = {3},
	issn = {2643-1564},
	url = {https://link.aps.org/doi/10.1103/PhysRevResearch.3.043207},
	doi = {10.1103/PhysRevResearch.3.043207},
	language = {en},
	number = {4},
	urldate = {2024-06-10},
	journal = {Physical Review Research},
	author = {Li, Zongda and Xu, Yiqing and Todd, Caleb and Xu, Gang and Coen, Stéphane and Murdoch, Stuart G. and Erkintalo, Miro},
	month = dec,
	year = {2021},
	pages = {043207},
}

@article{li_experimental_2020,
	title = {Experimental observations of bright dissipative cavity solitons and their collapsed snaking in a {Kerr} resonator with normal dispersion driving},
	volume = {7},
	issn = {2334-2536},
	url = {https://opg.optica.org/abstract.cfm?URI=optica-7-9-1195},
	doi = {10.1364/OPTICA.400646},
	language = {en},
	number = {9},
	urldate = {2024-06-10},
	journal = {Optica},
	author = {Li, Zongda and Xu, Yiqing and Coen, Stéphane and Murdoch, Stuart G. and Erkintalo, Miro},
	month = sep,
	year = {2020},
	pages = {1195},
}

@article{erdogan_cladding-mode_1997,
	title = {Cladding-mode resonances in short- and long-period fiber grating filters},
	volume = {14},
	issn = {1084-7529, 1520-8532},
	url = {https://opg.optica.org/abstract.cfm?URI=josaa-14-8-1760},
	doi = {10.1364/JOSAA.14.001760},
	language = {en},
	number = {8},
	urldate = {2024-06-10},
	journal = {Journal of the Optical Society of America A},
	author = {Erdogan, Turan},
	month = aug,
	year = {1997},
	pages = {1760},
}

@article{hoffman_ultrahigh_2014,
	title = {Ultrahigh transmission optical nanofibers},
	volume = {4},
	issn = {2158-3226},
	url = {https://pubs.aip.org/adv/article/4/6/067124/20832/Ultrahigh-transmission-optical-nanofibers},
	doi = {10.1063/1.4879799},
	language = {en},
	number = {6},
	urldate = {2024-06-10},
	journal = {AIP Advances},
	author = {Hoffman, J. E. and Ravets, S. and Grover, J. A. and Solano, P. and Kordell, P. R. and Wong-Campos, J. D. and Orozco, L. A. and Rolston, S. L.},
	month = jun,
	year = {2014},
	pages = {067124},
}

@article{coen_experimental_1998,
	title = {Experimental investigation of the dynamics of a stabilized nonlinear fiber ring resonator},
	volume = {15},
	issn = {0740-3224, 1520-8540},
	url = {https://opg.optica.org/abstract.cfm?URI=josab-15-8-2283},
	doi = {10.1364/JOSAB.15.002283},
	language = {en},
	number = {8},
	urldate = {2024-06-10},
	journal = {Journal of the Optical Society of America B},
	author = {Coen, S. and Haelterman, M. and Emplit, Ph. and Delage, L. and Simohamed, L. M. and Reynaud, F.},
	month = aug,
	year = {1998},
	pages = {2283},
}

@article{xue_microresonator_2017,
	title = {Microresonator {Kerr} frequency combs with high conversion efficiency},
	volume = {11},
	issn = {1863-8880, 1863-8899},
	url = {https://onlinelibrary.wiley.com/doi/10.1002/lpor.201600276},
	doi = {10.1002/lpor.201600276},
	language = {en},
	number = {1},
	urldate = {2024-06-10},
	journal = {Laser \& Photonics Reviews},
	author = {Xue, Xiaoxiao and Wang, Pei‐Hsun and Xuan, Yi and Qi, Minghao and Weiner, Andrew M.},
	month = jan,
	year = {2017},
	pages = {1600276},
}

\end{document}